\documentclass[aps,prb,reprint,groupedaddress,showpacs,amsmath,amssymb,twocolumn,floatfix]{revtex4-1}

\usepackage[T1]{fontenc}
\usepackage[latin9]{inputenc}
\usepackage{color}
\usepackage{units}
\usepackage{amssymb}
\usepackage{graphicx}
\usepackage{esint}
\usepackage{bm}
\usepackage{natbib}
\usepackage[breaklinks=true,colorlinks=true,urlcolor=blue,
citecolor=blue,linkcolor=blue,bookmarks=false]{hyperref}
\usepackage{setspace}
\usepackage{ulem}

\setcitestyle{journalcolor= blue}

\usepackage{babel}

\begin{document}
%%%%%%%%%%%%%%%%%%%%%%%%%%%%   Title and Author
\title{Enhanced electron-phonon coupling in doubly aligned hexagonal boron nitride bilayer graphene heterostructure}

\author{Manabendra Kuiri$^{1}$}
\author{Saurabh Kumar Srivastav$^{1}$}
\author{Sujay Ray$^{1}$}
\author{Kenji Watanabe$^{2}$}
\author{Takashi Taniguchi$^{2}$}
\author{Tanmoy Das$^{1}$}
\author{Anindya Das$^{1}$}
\affiliation{$^{1}$Department of Physics, Indian Institute of Science, Bangalore 560012, India}
\affiliation{$^{2}$National Institute of Material Science, 1-1 Namiki, Tsukuba 305-0044, Japan}
\email{anindya@iisc.ac.in}

\begin{abstract}
	
The relative twist angle in heterostructures of two-dimensional (2D) materials with similar lattice constants result in a dramatic alteration of the electronic properties. Here, we investigate the electrical and magnetotransport properties in bilayer graphene (BLG) encapsulated between two hexagonal boron nitride (hBN) crystals, where the top and bottom hBN are rotationally aligned with bilayer graphene  with a twist angle $\theta_t\sim 0^{\circ} \text{and}~ \theta_b < 1^{\circ}$, respectively. This results in the formation of two moiré superlattices, with the appearance of satellite resistivity peaks at carrier densities $n_{s1}$ and $n_{s2}$, in both hole and electron doped regions, together with the resistivity peak at zero carrier density. Furthermore, we measure the temperature(T) dependence of the resistivity ($\rho$). The resistivity shows a linear increment with temperature within the range 10K to 50K for the density regime $n_{s1} <n<n_{s2}$ with a large slope d$\rho$/dT $\sim$ 8.5~$\Omega$/K. The large slope of d$\rho$/dT is attributed to the enhanced electron-phonon coupling arising due to the suppression of Fermi velocity in the reconstructed minibands, which was theoretically predicted, recently in doubly aligned graphene with top and bottom hBN. Our result establishes the uniqueness of doubly aligned moire system to tune the strength of electron-phonon coupling and to modify the electronic properties of multilayered heterostructures.
%\textcolor{red}{interaction}. \sout{hello}

\end{abstract}

\keywords{Weyl semimetal, phase transition}

\maketitle
\section{Introduction}
Rotational alignment between atomically thin two-dimensional (2D) crystals leads to artificial superlattice potential forming moiré patterns\cite{novoselov20162d}. One well established example is graphene on hexagonal boron nitride (hBN). 
The weak periodic potential due to underlying hBN gives rise to modulation of graphene electronic band structure \cite{yankowitz2012emergence,yankowitz2019van,kumar2018localization}, with the emergence of clone Dirac cones\cite{hunt2013massive}, wherein, the Fermi velocity could be controlled by the relative twist angle between graphene and hBN, showing interesting physics like the Hofstader butterfly \citep{dean2013hofstadter}, resonant tunneling\cite{mishchenko2014twist}, change of topological winding number\cite{wang2015topological}, topological valley current\cite{endo2019topological}, Brown-Zak oscillations\cite{ponomarenko2013cloning}etc. Moreover, twisting individual layers leads to flat bands \cite{bistritzer2011moire}, where emergent phenomenon like  correlated insulating state\cite{cao2018correlated}, superconductivity\cite{cao2018unconventional}, quantum anomalous Hall effect\cite{serlin2019intrinsic}, fractional Chern insulating states\cite{chen2020tunable,repellin2019chern}, moiré excitons\cite{tran2019evidence},  and ferromagnetism\cite{serlin2019intrinsic} has been observed. Furthermore, in multilayered heterostructures, competing moiré superlattice could be present, due to the lattice mismatch between individual layers which can lead to further dramatic change in the electronic properties\cite{wang2019new,wang2019composite}. \\

%%%%%%%%%%%%%%%%%%  FIGURE 1  %%%%%%%%%%%%%%%%%%%%%%%%
\begin{figure*}[tbh]
	\includegraphics[width=0.95\textwidth]{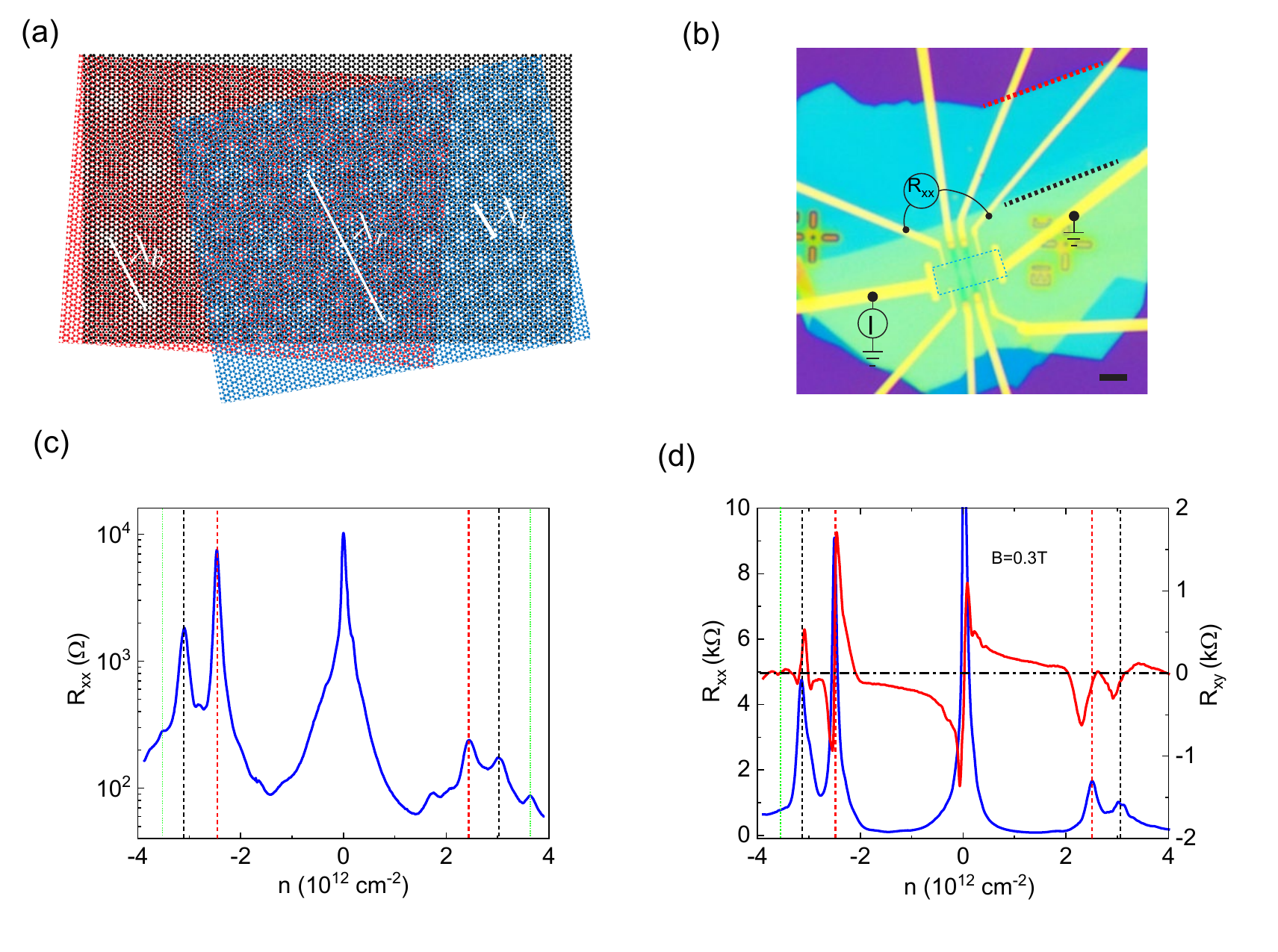}
	\caption{ (a) Schematic illustration of the moiré superlattice. Bilayer graphene (black) is encapsulated between the top (bottom) hBN marked with blue (red), respectively. (b) Optical image of the device with the measurement scheme. The edge of the top (bottom) hBN are marked with black (red) dashed line. Scale bar $\sim$2$\mu$m. (c) Four terminal resistance ($R_{xx}$) as a function of carrier density ($n$) measured at T$\sim 250$ mK. The red(black) vertical dashed lines corresponds to SDP1 (SDP2), and the green dotted line corresponds to one of the supermoiré peaks. (d) Hall resistance $R_{xy}$ and longitudinal resistance $R_{xx}$, as a function of carrier density for B=0.3T. The Hall resistance changes sign at PDP, SDP1 and SDP2 indicating that the Fermi energy crosses the superlattice band. Similarly, Hall resistance also changes sign near  $n\sim \pm 3.6\times 10^{12}~\text{cm}^{-2}$, which correspond to one of the supermoiré wavelengths.}
	\label{fig:fig1}
\end{figure*}
%%%%%%%%%%%%%%%%%%%%%%%%%%%%%%%%%%%%%%%%%%%%%%%%%%%%%%%%%%%%%%%%%%%%%%%%%

%%%%%%%%%%%%%%%%%%  FIGURE 2 %%%%%%%%%%%%%%%%%%%%%%%%%%%%%%%%%%
\begin{figure*}[tbh]
	\includegraphics[width=0.9\textwidth]{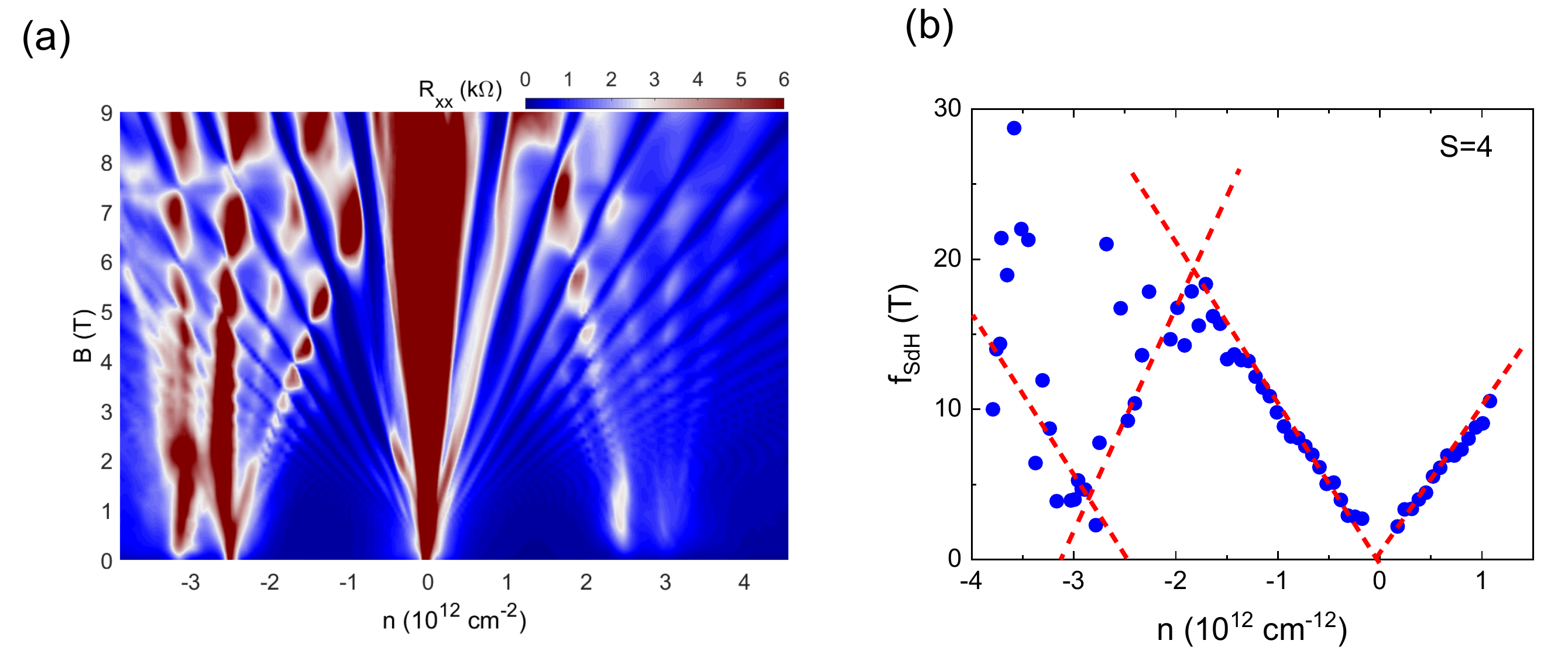}
	\caption{(a) Landau fan diagram as a function perpendicular magnetic field (B), and carrier density ($n$) at T$\sim$250mK. Quantum oscillations are observed at PDP with $\nu=4,8,12....$ and also with SDP1 and SDP2 with fillings $\nu=4$ evolving from $n_{s1}$. (b) Extracted SdH oscillation frequency as a function of  carrier density obtained from (a). }
	\label{fig:fig2}
\end{figure*}

Alternatively, another concomitant way to engineer moiré superlattice is to exploit the rotational alignment of graphene with both top and bottom hBN. This could result in two moiré superlattice structures of wavelength $\lambda_t$ and $\lambda_b$, respectively. The interference of these two moiré superlattices could result in a supermoiré structure of wavelength $\lambda_r$. The supermoiré wavelength $\lambda_r$ can vary from $\sim$1~nm to infinity, depending on the rotational alignment of graphene with top (bottom) hBN\cite{wang2019new}. Recently, the transport measurements have been carried out for hBN/graphene/hBN doubly aligned heterostructures and the signature of supermoiré patterns have been observed\cite{wang2019composite}. Furthermore, in ref\cite{andjelk2020double} it has been theoretically predicted the band is less dispersive with reduced Fermi velocity for the minibands in doubly aligned hBN/graphene/hBN heterostructures. Since the BLG has a parabolic dispersion, it is expected to exhibit a further reduction in Fermi velocity in doubly aligned hBN/BLG/hBN heterostructures as compared to the monolayer counterpart. However, there are no experimental studies on this kind of structures.
 
Here, we study the low temperature electrical transport in a hBN/BLG/hBN heterostructure, where the top and bottom hBN are rotationally aligned with BLG  with a twist angle $\theta_t\sim 0^{\circ} \text{and} ~\theta_b < 1^{\circ}$, respectively.  The low-temperature transport shows the emergence of additional  satellite resistivity peaks at carrier densities ($n_{s1}, n_{s2}$) in hole as well as electron doped regions, together with the intrinsic resistivity peak at zero carrier density. The Hall resistance at low magnetic field shows expected sign change at these peak positions. Furthermore, we study the temperature dependence of resistivity as a function of carrier density. We find that resistivity ($\rho$) varies with temperature as $\sim \rho_0(n)+AT^\beta$ for different carrier densities. Interestingly,  we find that, in the density regime between $n_{s1} <n<n_{s2}$,  the measured resistivity varies linearly ($\beta\sim1$) with temperature within the range 10K to 50K. The slope of resistivity (d$\rho$/dT) in this linear regime was found to be maximum of $\sim 8.5~\Omega/$K. This value is approximately two orders of magnitude higher than pristine graphene with d$\rho$/dT$~\sim 0.1~\Omega/$K\cite{chen2008intrinsic,PhysRevLett.105.256805,PhysRevB.77.115449}. The linear dependence of resistivity with temperature arises due to the electron-acoustic phonon scattering\cite{PhysRevB.77.115449,wu2019phonon}. It was recently shown theoretically in ref\cite{andjelk2020double}, that in doubly aligned hBN/graphene/hBN heterostructures, the minibands are less dispersive with reduced Fermi velocity in the density regime $n_{s1} <n<n_{s2}$. In our device, we attribute the higher value of d$\rho$/dT to the enhanced electron-phonon coupling resulting from the suppression of Fermi velocity in these minibands.

%% Further extensive experimentally and theoretical work is desirable to explore the band structure reconstruction and the possibility of appearance of flat electronic dispersion\cite{finney2019tunable,andjelk2020double}.

\section{Results and Discussion}

We fabricated bilayer graphene (BLG) devices encapsulated between two hexagonal boron nitride crystals using the dry transfer technique\cite{wang2013one,PhysRevB.98.035418,kuiri2019energetics}. The fabrication technique is similar to our earlier work\cite{kumar2016tunability,kumar2018localization,kuiri2019energetics}. In addition, the BLG edge was carefully aligned with the
crystallographic axes of both top (bottom) hBN. The natural rectangular shape of bilayer graphene allowed us to pattern the device into a Hall bar geometry. Edge contacts were established by standard electron beam lithography followed by thermal deposition of Cr/Pd/Au (2/10/70)nm. The optical image of the device is shown in Fig.1(b). The measurements were carried out in $^3$He cryostat using standard lock-in technique, in a four-terminal configuration as schematically shown in Fig.~1(b).\\

Figure 1(a) shows the illustration of the appearance of moiré pattern, where the two crystals, BLG and hBN are aligned relative to one another ($\theta_t$, with top hBN) and ($\theta_b$, with bottom hBN). The relative rotation between the two crystals defines the moiré wavelength $\lambda_i$. The appearance of supermoiré structure has been illustrated in Fig.~1(a). In our device the top hBN is nearly perfectly aligned with bilayer graphene ($\theta_t\sim 0^{\circ}$), whereas the bottom hBN has slightly different twist angle ($\theta_b \leq 1^{\circ}$). Fig. 1(c) shows the four terminal longitudinal resistance as a function of carrier density ($n$) at T$\sim$ 250mK. The estimated mobility of our device was found to be $\mu\sim 57000~\text{cm}^2$/Vs(Supplemental Material\cite{supplementary}). In addition to the resistance peak at primary Dirac point (PDP), we also observe two strong resistance peak at $n_{s1}\sim -2.46\times 10^{12}~\text{cm}^{-2}$ and $n_{s2}\sim -3.09\times 10^{12}~\text{cm}^{-2}$, highlighted by red and black dashed vertical line in Fig. 1(c), respectively. The resistance peaks corresponding to densities $n_{s1}$ and $n_{s2}$ originate due to the two different moiré superlattice potentials arising from the rotational alignment of top and bottom hBN with BLG. We attribute the satellite resistance peaks corresponding to densities $n_{s1}$, $n_{s2}$   as secondary Dirac point 1 (SDP1) and secondary Dirac point 2 (SDP2), respectively. It was reported that the second satellite peak (SDP2) may appear even in single aligned graphene/hBN heterostructure due to formation of Kekule superstructure with $n_{s2}\sim\text{1.65}n_{s1}\cite{chen2017emergence}$. However, for our device $n_{s2}\sim \text{1.2}n_{s1}$. Thus, we rule out the origin of SDP2 due to the formation of Kekule superstructure. It is worth to mention here that the spatial distribution of the twist angles over the sample should be uniform to observe the strong side resistivity peaks.

%%%%%%%%%%%%%%%%%%%  FIGURE 3  %%%%%%%%%%%%%%%%%%%%%%%%%%%
\begin{figure*}[tbh]
	\includegraphics[width=1\textwidth]{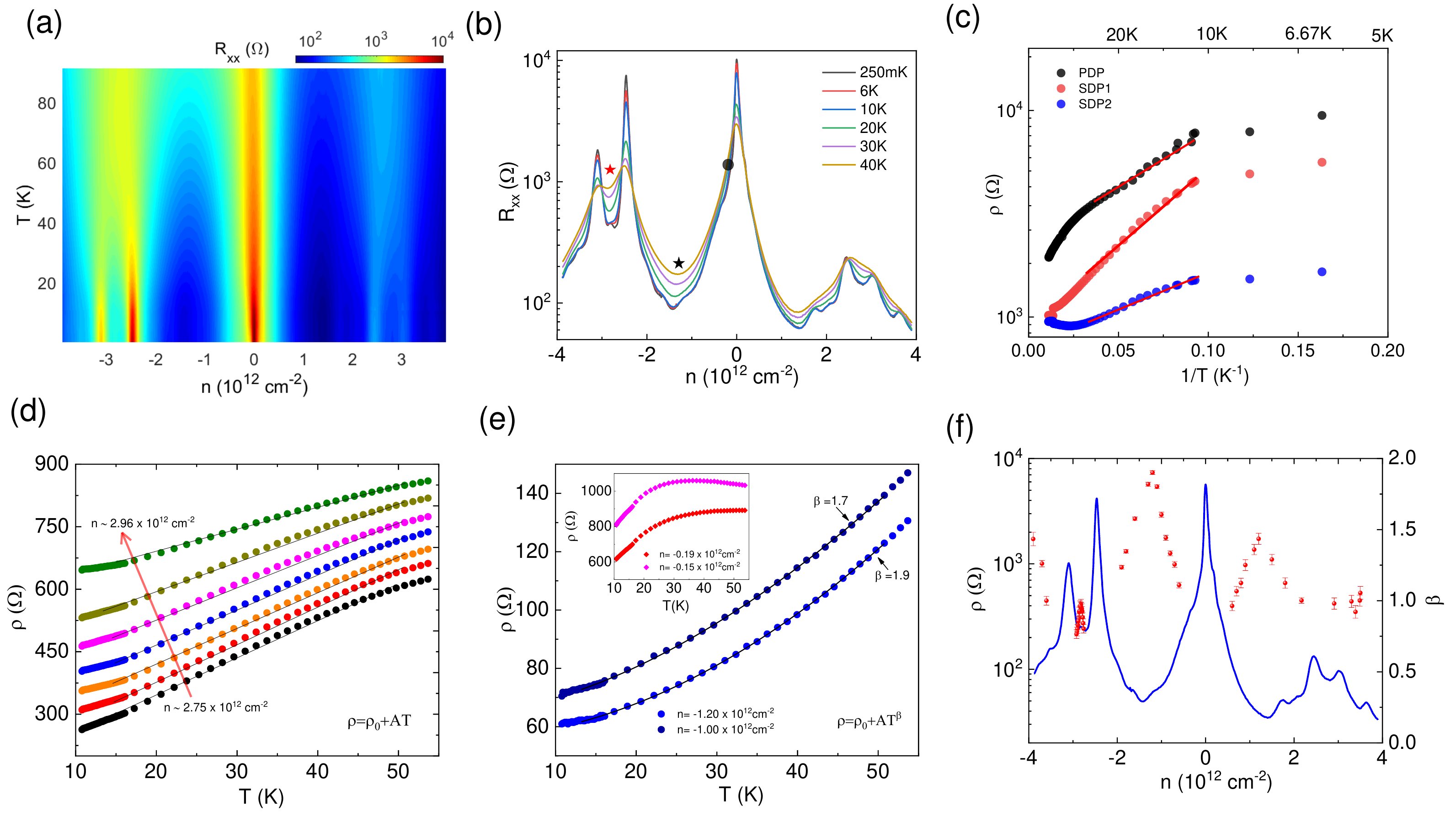}
	\caption{(a) Four terminal resistance as a function of carrier density, for several values of temperature from T$\sim$250mK to 90K. (b) Line traces of resistance vs carrier density from (a) for T$\sim$250mK, 10K, 30K and 40K. (c) Arrhenius plot of resistivity $\rho$ vs 1/T at PDP, SDP1 and SDP2 showing activated transport behavior, yielding a gap, $\Delta_{SDP1}\sim 3$meV and $\Delta_{SDP2}\sim 1.2 $meV. The dashed line shows the linear fit. (d) $\rho$ vs T for several values of $n$ ranging in between $n_{s1} <n<n_{s2}$, highlighted as red star in (b). The solid lines shows the fit with $\rho\sim \rho_0(n)+AT$, for the temperature range for 10K to 50K, with a slope of A=d$\rho$/dT$\sim 8.5~\Omega/\text{K}$. The curves are offset by 50$\Omega$ for clarity. (e) $\rho$ with T for few other values of carrier densities highlighted as black star in (b). The solid lines are the fit with $\rho\sim \rho_0(n)+AT^\beta$. Inset shows the $\rho$ vs T for the region marked as black circle in (b). (f) $\rho$ and $\beta$ as a function of carrier density. }
	\label{fig:fig3}
\end{figure*}
%%%%%%%%%%%%%%%%%%%%%%%%%%%%%%%%%%%%%%%%%%%%%%%%%%%%%%%%%%%%%%%%%%%%%%%%%

In contrast to the strong resistance peak in the hole doped regime, weak resistance peaks are observed in the electron doped regime, at similar densities. This asymmetry in electron-hole transport in BLG/hBN moiré devices arises because the hBN induces different onsite energies, which is stronger for hole band as compared to the electron band\cite{yankowitz2012emergence,PhysRevB.91.245422,andjelk2020double}. The electron-hole asymmetry has also been observed experimentally\cite{yankowitz2012emergence,wang2019composite}  and calculated theoretically\cite{andjelk2020double} for monolayer graphene aligned with two hBN, citing a similar origin. Now, from the carrier density of the satellite peak, we can estimate  moiré wavelength ($\lambda_i$) given by $n_{s}=8/\sqrt 3 \lambda_i ^2$. The resistance peak corresponding to carrier density $n_{s1}$ shows a perfect alignment of BLG with top hBN ($\theta_t\sim 0.0^{\circ}$) with a moiré wavelength of $\lambda_t \sim 13.7$nm. Similarly, the resistance peak at $n_{s2}$ corresponds to a moiré  wavelength of $\lambda_b\sim 12.2$nm ($\theta_b\sim 0.56^{\circ}$). The interference of these two moiré superlattices could  result in the formation of supermoiré structure. In ref\cite{andjelk2020double}, the supermoiré wavelengths have been theoretically calculated based on four possible reciprocal lattice vectors configurations. For our device $\theta_t\sim 0.0^{\circ}$ and $\theta_b\sim 0.56^{\circ}$, the possible supermoiré wavelengths are $\sim$ 25.6, 23.4, 11.7 and 9.3nm(Supplemental Material\cite{supplementary}). The resistance peaks observed in our experiment at  $n\sim \pm 3.6\times 10^{12}~\text{cm}^{-2}$(marked as green dotted line in Fig.1c) corresponds to moiré wavelength of $\sim$11.5nm, which matches closely with one of the supermoiré wavelength (11.7nm). However, weak shoulder at $n\sim \pm 1.9\times 10^{12}~\text{cm}^{-2}$ does not match with any supermoiré wavelength. In ref\cite{wang2019composite}, it was proposed that these peaks could be due to more exotic superlattice phenomena or higher order moiré periodicities. Fig. 1(d) shows the Hall resistance $R_{xy}$ as a function of carrier density for a perpendicular magnetic field of B=0.3T. As one would expect the charge carriers change from hole-like to electron like when crossing both the charge neutrality $n=0$, and at superlattice gaps for $n_{s1}$, and $n_{s2}$. We would also like to note that, the Hall resistance changes sign at $n\sim - 3.6\times 10^{12}~\text{cm}^{-2}$, which corresponds to the supermoiré wavelength of $\sim$11.7nm, highlighted by green dotted vertical line in Fig.~1(d).

\section{Magnetic field data}

We measure the Hall resistance, $R_{xy}$ and the longitudinal resistance, $R_{xx}$ as a function of a perpendicular magnetic field (B) and carrier density ($n$). We observe quantum oscillations evolving from PDP, SDP1, and SDP2 for both electron and hole side. Fig. 2(a) shows the measured $R_{xx}$ as a function of carrier density and magnetic field for both electron and hole-doped regions. The sequence of $R_{xy}$ verifies that indeed our flake is bilayer graphene (Supplemental Material\cite{supplementary}).
%The gate voltage axis in Fig. 1(c) was converted to the carrier density per moire unit cell ($n/n_0$). $V_{bg}=-35V$ corresponds to SDP, resulting a carrier of  density $n\sim - 2.54\times 10^{12}cm^{-2}$.  We define $4n_0$ as the density required to fill the mini Brillouin zone. The prefactor 4 arises from spin and valley degeneracies. This density gives a twist angle of $\theta_t\approx 1.02^{\circ}$ (Supplemental material Sec. III for details). In the subsequent section, all data will be presented in terms of normalized carrier density ($n/n_0$). In presence of perpendicular magnetic field B, gaps evolve in the energy spectrum of moire BLG, which follows Diophantine equations\cite{dean2013hofstadter}
%\begin{equation}
%\frac{n}{n_0}=\nu\frac{\phi}{\phi_0}+s
%\end{equation}
%where, $\phi_0=h/e$ represents the flux quantum. For $s=0$, $\nu$ corresponds to LL filling factors for intrinsic bilayer graphene. Thus tracing the minima in $R_{xx}$ as a function of $n/n_0$ and magnetic field will give the filling factor $\nu$ for a fixed $s$. $s\neq0$ corresponds to the moire band. It can be clearly seen that Landau fan evolves from $n/n_0=0, \pm4, \pm5$ both electron and hole side as shown in Fig. 2(a). The crossing pattern for $n/n_0$ =-5 to -6,represents Brown Zak oscillations\cite{ponomarenko2013cloning}.
Alternatively, we extract the Shubnikov de Haas (SdH) oscillation frequency ($f_{SdH}$), from which we can calculate the Landau level(LL) degeneracy for $n=0$ and for $n_{s1}, n_{s2}$\cite{cao2018correlated}. The SdH oscillation frequency is given by $f_{SdH}=\phi_0|n|/S$, where,  S represents the degeneracy of the Landau levels. In BLG, due to the spin and valley degeneracy S=4. Fig. 2(b) shows the measured oscillation frequency as a function of carrier density. Near PDP, we obtain $S=4$, revealing the observed fillings at $\nu=\pm 4,\pm8,\pm12..$, consistent with the LL spectrum of intrinsic bilayer graphene. Similarly, the LL fan for $n_{s1}$ and $n_{s2}$ also has a degeneracy  of S=4.

%The resistance peaks for $n\sim\ptm 2.54\times 10^{12} cm^{-2}$ corresponds to the secondary Dirac point (SDP), and that corresponding to $n\sim\pm 3\times 10^{12} cm^{-2}$, defines the tertiary Dirac point (TDP)  We define density $n_0=4/A$, where A is the moire unit cell area. as fully filled MoIRE band, i.e no of electrons required to fill electrons a Moire band corresponding to fully filled band at $n\sim\pm 2.54\times 10^{12} cm^{-2}$. We also observe additional resistivity peaks at the electron side at $n\sim 3.5\times 10^{12} cm^{-2}$, with a filling of $n_0=6$.\\

\section{Temperature Dependence}

The dependence of material's resistivity with temperature reveals important physics related to carrier scattering, electron-phonon coupling, and electron-electron interactions in the system\cite{hartnoll2015theory}. Fig. 3(a) shows the 2D colorplot of  $R_{xx}(T)$, as a function of carrier density ($n$). Insulating behavior was observed at PDP, SDP1, and SDP2. Fig. 3(b) shows the line cut of Fig. 3(a) at several temperatures. It can be seen that the resistivity increases with the increase in temperature close to $n\sim -2.8\times 10^{12} \text{cm}^{-2}$ in between SDP1, and SDP2 signifying metallic transport. Fig.~3(c) shows the Arrhenius plot $\rho(T)$ vs 1/T at PDP, SDP1, and SDP2. Activated transport is observed for PDP and SDP1. However, although the SDP2 shows activated behavior at low temperature, the resistivity increases at high temperatures. The linear fit with the Arrhenius equation gives a gap of $\Delta_{ns1}\sim 3~$meV and $\Delta_{ns2}\sim 1.2~$meV for  SDP1, and SDP2, respectively. In Fig. 3(d), we show $\rho$ with T, for several values of $n$ between $n_{s1}$ and $n_{s2}$. We see that $\rho$ evolves linearly with T up to T$\sim$ 50K. We fit our temperature dependence data in the range of 10K to 50K with $\rho(n,T)\sim \rho_0(n)+AT^{\beta}$. We find that in the region between $n_{s1} <n<n_{s2}$,  $\rho(n,T)\sim \rho_0+AT$ with $\beta\sim 1$, and a maximum value of (A) is found to be A=d$\rho$/dT$\sim 8.5~\Omega/$K. In Fig.~3(e) we show the $\rho$ -T for few other densities. It can be seen in Fig.~3(e), that in the region between PDP and SDP1, $\beta$ reaches upto $\sim$1.9. Close to PDP, resistivity increases non monotonically with temperature, as can be seen in the inset of Fig. 3e. 

 The value of d$\rho$/dT in the regime $n_{s1} <n<n_{s2}$ is nearly two orders of magnitude larger as compared to intrinsic monolayer graphene with d$\rho$/dT$\sim 0.1 \Omega/$K\cite{chen2008intrinsic,PhysRevLett.105.256805,PhysRevB.77.115449}. It is worth to mention here that the resistivity of intrinsic bilayer graphene shows a very weak dependence on temperature\cite{dean2010boron,PhysRevB.82.081409}.In Fig.~3(f), we plot $\beta$ as a function of carrier density. It can be seen that $\beta\sim 1$, in the region between $n_{s1} <n<n_{s2}$ in the hole doped region. Qualitatively, similar behavior was observed in electron doped regime with a smaller value of d$\rho$/dT$\sim 0.9~\Omega/$K (Supplemental material\cite{supplementary}). However, at all other densities $\beta$, changes significantly with the change in carrier densities as can be seen in Fig. 3(f). Recently, it has been shown theoretically\cite{andjelk2020double}, that in the doubly aligned hBN/graphene/hBN heterostructures, reconstruction of
	band structures gives rise to the formation of less
	dispersive minibands with reduced Fermi velocity. Now
	having BLG in our device over mono layer, could
	imply further reduction in Fermi velocity since the low
	energy dispersion is parabolic in BLG. Thus one expects
	further reduction in Fermi velocity for hBN/BLG/hBN
	heterostructures. This could give rise to large electron-phonon coupling similar to what has been observed in twisted bilayer graphene devices\cite{polshyn2019large}. Due to this enhanced electron-phonon coupling, we observe the large d$\rho$/dT of $\sim 8.5\Omega/$K  in the density regime
	$n_{s1} <n<n_{s2}$. To compare our results with mono layer
	graphene, we follow the similar approach, which was used
	for monolayer graphene and twisted bilayer graphene. Theoretically, the resistivity at high T, due to acoustic phonon induced scattering is given by\cite{PhysRevB.77.115449,wu2019phonon}
\begin{eqnarray}
\rho(T)=\frac{\pi D^2}{g_s g_v e^2 \hbar \rho_m v_F^2 v_{ph}^2} k_B T
\end{eqnarray}
Where D, $v_{ph}$, $v_F$, and $\rho_m$ are the deformation potential, phonon velocity, Fermi velocity of graphene, and atomic mass density, respectively. $g_s(g_v)$ are the spin(valley) degeneracy. For pristine graphene $\rho_m=7.6\times 10^{-7} \text{kg/m}^2$, $v_{ph}=2\times 10^4 \text{m/s}$, $v_F=10^6 \text{m/s}$, and $D\sim 20$eV yields d$\rho$/dT$\sim 0.1 \Omega/$K\cite{chen2008intrinsic,PhysRevLett.105.256805,PhysRevB.77.115449}. Since the phonon spectrum remains relatively invariant in moiré system\cite{PhysRevB.98.241412,wu2019phonon}, $v_{ph}$ is assumed to be constant. Therefore, using the above values for $D$, $\rho_m$, $v_{ph}$ and the experimentally measured value of d$\rho$/dT$\sim 8.5 \Omega/$K, we estimate the renormalized Fermi velocity $v_F\sim0.1\times 10^6$ m/s. Please note that, different values of $D\sim (10-30)~$eV are quoted in theory\cite{PhysRevB.77.115449} and even this uncertainty in $D$ can not explain the nearly two order of magnitude enhancement in d$\rho$/dT. We would also like to mention that temperature broadening and the excitation of carriers across the minigaps could affect the temperature dependence of resistivity. However, in our device, we believe that this would not affect significantly due to following reason. The SDP1 and SDP2 peaks are separated by an energy interval of $\sim$25 meV(Supplemental Material\cite{supplementary}). Since, we have performed our analysis near the middle of the two secondary peaks, which is $\sim$12 meV away from SDP1 and SDP2. The estimated width of the SDP1 and SDP2 was found to be $\sim$1meV and $\sim$1.5meV, respectively (Supplemental Material\cite{supplementary}), which is less compared to the temperature broadening of $\sim$4.3 meV at 50K. This suggests that the temperature broadening effect will be dominant only near the SDP1 and SDP2 in energy window of $\sim$4meV, and it will not affect significantly near the density regime, where we have performed our analysis. Therefore, we attribute the large d$\rho$/dT to the suppression of $v_F$ in the reconstructed bands, as a plausible expansion. This reduced Fermi velocity, in our device, is not surprising. As discussed earlier it could be due to the formation of narrow bands which has been theoretically predicted in doubly aligned graphene devices with hBN. Our result, therefore, hints on a perspective of creating an alternative framework for narrow bands using doubly aligned graphene on hBN devices. Further experimental and theoretical work is desirable to explore the band structure of doubly aligned bilayer graphene devices, and to explore the possibility of obtaining flat dispersion, which could be an interesting platform to study strong interaction physics or superconductivity as observed in twisted bilayer graphene\cite{cao2018unconventional}.

\section{Conclusion}
In summary, we have presented the electrical and magneto-transport properties of doubly aligned hBN/BLG/hBN moiré heterostructure. We observe the appearance of strong resistivity peaks due to the presence of two moiré wavelengths. We show that the temperature dependence of resistivity scales linearly with T in the region $n_{s1} <n<n_{s2}$, with slope of nearly two orders of magnitude larger than pristine graphene.
\section{Acknowledgement}
The authors would like to acknowledge the Center for Nanoscience and Engineering (CENSE), IISc for fabrication facilities. S.K.S. acknowledges PMRF, MHRD for financial support. A.D. thanks the Department of Science and Technology (DST), India for financial support (DSTO-2051) and acknowledges the Swarnajayanti Fellowship of the DST/SJF/PSA-03/2018-19. A.D. also acknowledges supports from the MHRD, Govt. of India under STARS research funding (STARS/APR2019/PS/156/FS).

%\bibliography{ref_blg}
%%%%%%%%% Supplemental material %%%%%%%%%%
\pagebreak
\widetext
\newpage
\begin{center}
	\textbf{\large Supplemental Material for Enhanced electron-phonon coupling in doubly aligned hexagonal boron nitride bilayer graphene heterostructure}
\end{center}

\begin{center}
	Manabendra Kuiri$^{1}$, Saurabh Kumar Srivastav$^{1}$, Sujay Ray$^{1}$, Kenji Watanabe$^{2}$, Takashi Taniguchi$^{2}$, Tanmoy Das$^{1}$, and Anindya Das$^{1}$
	
	\it{$^{1}$Department of Physics, Indian Institute of Science, Bangalore 560012, India}
	
	\it{$^{2}$National Institute of Material Science, 1-1 Namiki, Tsukuba 305-0044, Japan}
	
\end{center}

\setcounter{equation}{0}
\setcounter{figure}{0}
\setcounter{table}{0}
\setcounter{page}{1}
\makeatletter
\renewcommand{\theequation}{S\arabic{equation}}
\renewcommand{\thefigure}{S\arabic{figure}}
\renewcommand{\bibnumfmt}[1]{[S#1]}
\renewcommand{\citenumfont}[1]{S#1}

\section{Device characterization}
The measured four-terminal resistance ($R$) vs backgate voltage ($V_g$) was fitted with the equation

\begin{eqnarray}
	R=R_c+\frac{L}{We\mu\sqrt{n_0^2+n^2}}
\end{eqnarray}
where, $R_c$, $L$, $W$, $e$, and $\mu$ are the contact resistance, device length, device width, electronic charge and mobility, respectively. $n_0$ corresponds to the charge inhomogeneity. The carrier density ($n$) is given by $\frac{C_g (V_g-V_{d})}{e}$, where $C_g$, $V_d$ are the capacitance per unit area of the bottom gate and gate voltage at the charge neutrality point, respectively. Fig.~S1 shows the fitting with Eq.(S1) near the primary Dirac point. 

\begin{figure*}[tbh]
	\includegraphics[width=0.5\textwidth]{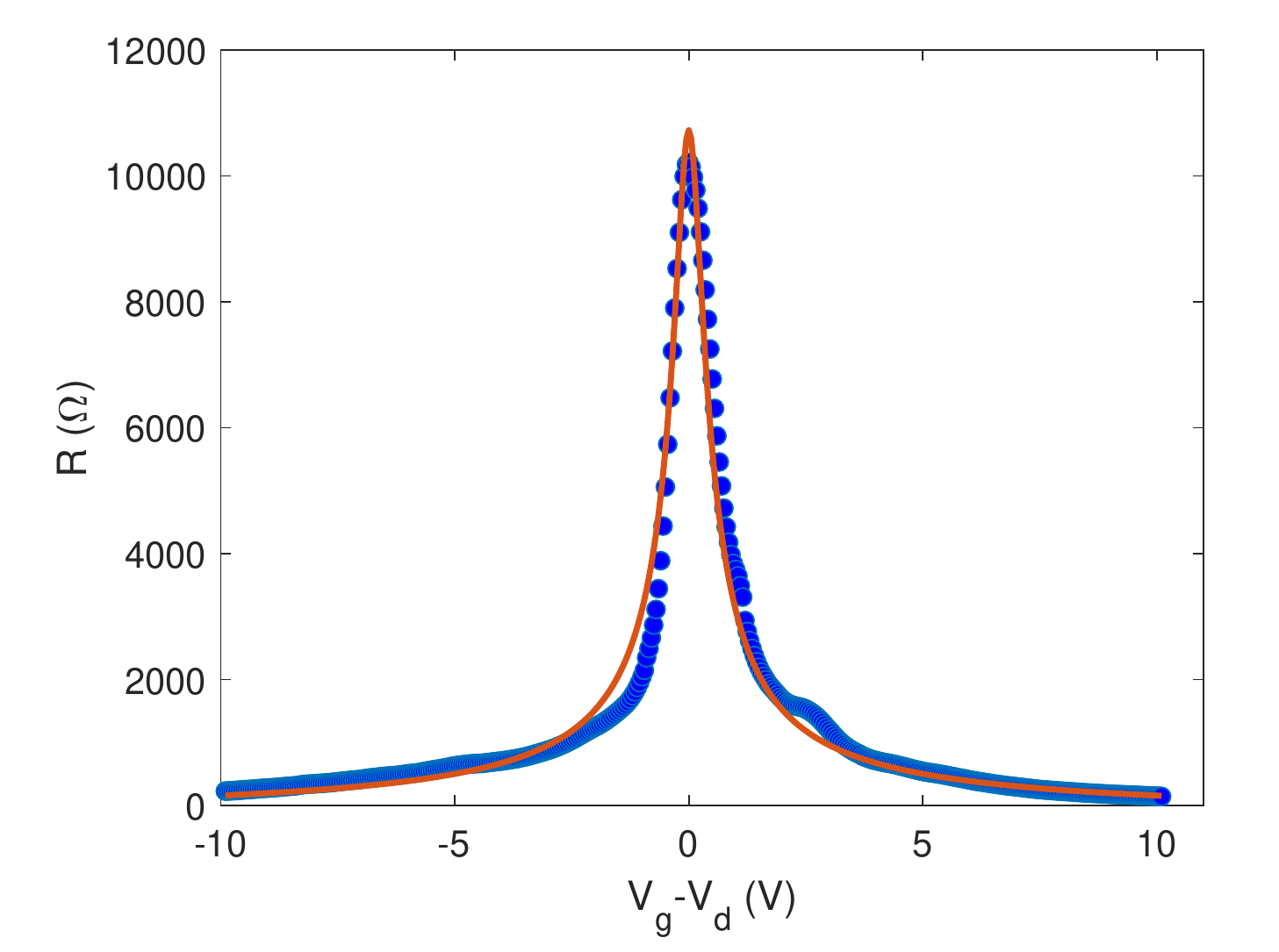}
	\caption{ Four terminal resistance as a function backgate voltage ($V_g$) at T$\sim$250mK. The blue filled circles show the experimentally measured data, and the red solid line shows the fit with Eq.(S1). The fit gives a mobility of $\mu\sim 57000~\text{cm}^2/\text{Vs}$. }
\end{figure*}
%%%%%%%%%%%%%%%%%%%%%%%%%%%%%%%%%%%%%%%%%%%%%%%%%%%%%%%%%%%%%%%%%%%%%%%%%

In Fig. S2, we estimate the charge inhomogeneity at SDP1 and SDP2 for the electron and the hole side. This was done by taking the log-log plot of conductance as a function of carrier density and extapolating the linear conductance on the $n$ axis. This gives carrier in-homogeneity $\delta n\sim (2-3.5)\times 10^{10}\text{cm}^{-2}$ at SDP1, and $\delta n\sim (4-4.5)\times 10^{10}\text{cm}^{-2}$ for SDP2, respectively. From this value of charge inhomogeneity, we estimate the  Fermi energy broadening at SDP1 and SDP2 to be $\Delta E_F\sim $1 meV, and $\Delta E_F\sim $1.5 meV, respectively. We used the relation $E=\hbar^2 k^2/2m^*$, where $m^*=0.033m_e$\cite{PhysRevB.82.081409}, $m_e$ being the electronic charge, and $k=\sqrt{\pi \delta n}$. The distance between the center of two peaks SDP1 and SDP2 in density axis is $\sim0.63\times 10^{12} \text{cm}^{-2}$, which translates to energy scale gives $\sim 25$meV.

\begin{figure*}[tbh]
	\includegraphics[width=0.8\textwidth]{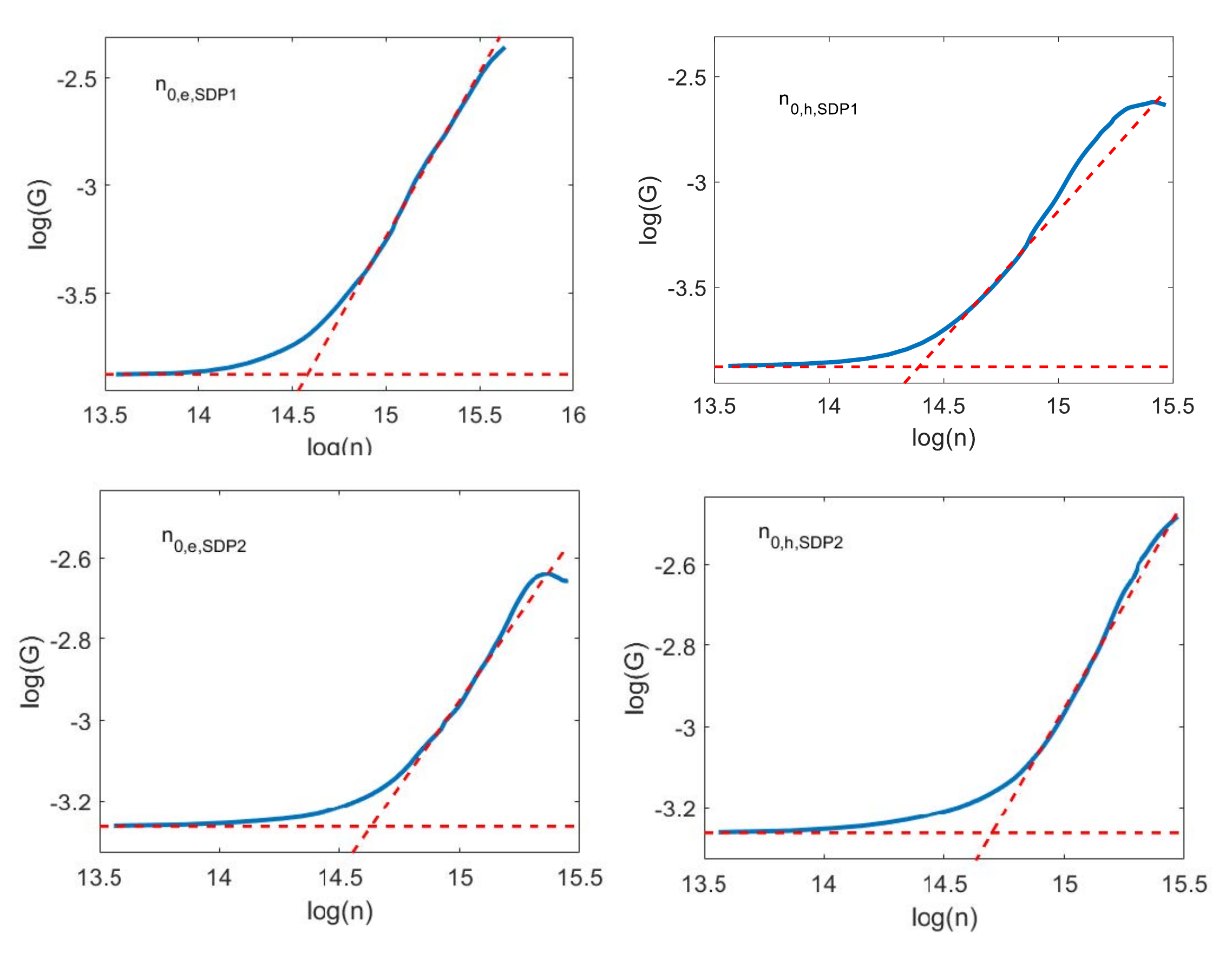}
	\caption{Log-Log plot of conductance G($e^2/h$) versus carrier density showing carrier in-homogeneity $\delta n\sim (2-3.5)\times 10^{10}\text{cm}^{-2}$ at SDP1, and $\delta n\sim 4\times 10^{10}\text{cm}^{-2}$ for SDP2. }
\end{figure*}
%%%%%%%%%%%%%%%%%%%%%%%%%%%%%%%%%%%%%%%%%%%%%%%%%%%%%%%%%%%%%%%%%%%%%%%%%

%\pagebreak
\newpage

\section{Quantum Hall Data}

%%%%%%%%%%%%%%%%%%  FIGURE 2 %%%%%%%%%%%%%%%%%%%%%%%%%%%%%%%%%%
\begin{figure*}[tbh]
	\includegraphics[width=0.9\textwidth]{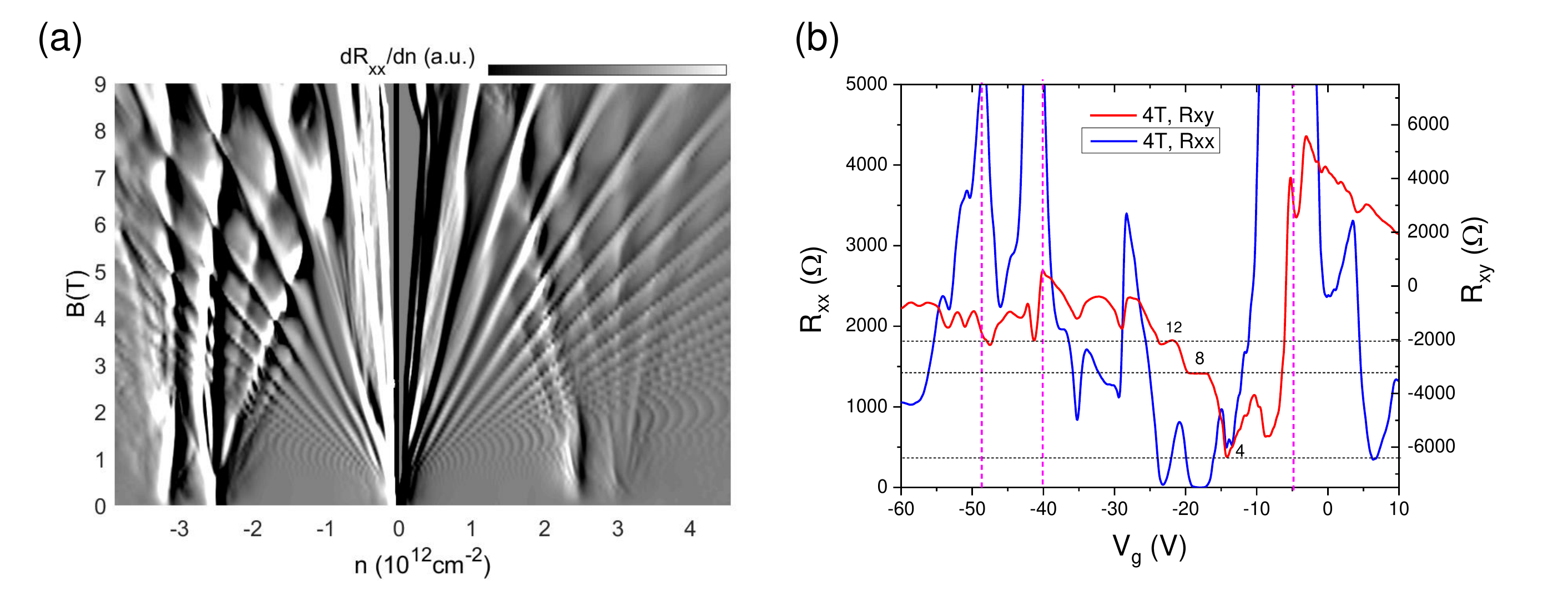}
	\caption{(a) The derivative of measured longitudinal resistance as a function of carrier density shown in Fig. 2a of the main text. (b) The Hall resistance at B=4T shows a sequence of  $\nu=4,8,12..$}
	%\label{fig:fig2}
\end{figure*}
%%%%%%%%%%%%%%%%%%%%%%%%%%%%%%%%%%%%%%%%%%%%%%%%%%%%%%%%%%%%%%%%%%%%%%%%%
\newpage

\section{Supermoire features}
The supermoir\'e  wavelength can be calculated using Ref\cite{andjelk2020double}. The unit cell vectors of
graphene are given as
\begin{eqnarray}
	\mathbf{a}^{gr}_{1}&=a[1,0], \\
	\mathbf{a}^{gr}_{2}&=a[1/2, \sqrt{3}/2], 
\end{eqnarray}

and the unit vectors for the bottom ($i=1$) and top ($i=2$) hBN are given by
\begin{equation}
	\mathbf{a}^{hBN_i}_{1,2}=\mathbf{R}(\theta_i)\mathbf{a}^{gr}_{1,2}/(1+\delta_{i}),
\end{equation} 
where $a\approx 0.246$ nm is lattice constant for graphene. $\mathbf{R}(\theta_i)$ is defined as the rotation matrix in anticlockwise direction with angle $\theta_i$, and $\delta_{1,2}$ being the lattice mismatch between graphene and top and bottom hBN lattice constants, with $\delta\approx 0.018$. Now, the Reciprocal vectors are given by $\mathbf{b}^{gr}_{1}=2\pi/a[1, -\sqrt{3}/3]$, $\mathbf{b}^{gr}_{2}=2\pi/a[0, 2\sqrt{3}/3]$, $\mathbf{b}^{hBN_i}_{1,2}=\mathbf{R}(\theta_i)\mathbf{b}^{gr}_{1,2}/(1+\delta_{i})$ satisfying $\mathbf{b}^{gr}_j \mathbf{a}^{gr}_k=\mathbf{b}^{hBN_i}_j \mathbf{a}^{hBN_i}_k=2\pi\delta_{j,k}$.

The supermoir\'e vector can be defined as

\begin{eqnarray}
	\mathbf{b}^{SM}=i\mathbf{b_1}^{M_1}+j\mathbf{b_2}^{M_1}-k\mathbf{b_1}^{M_2}-l\mathbf{b_2}^{M_2}
\end{eqnarray}
where, $i,j,k,\text{and}~l$ are intergers.
In our doubly aligned hBN/BLG/hBN heterostructure, under the condition $\theta_t\sim 0^{\circ}$, and $\theta_b\sim 0.56^{\circ}$, four possible supermoir\'e wavelengths are given by  $\lambda_r^{SM}$

\begin{equation}
	\begin{split}
		\lambda_1^{SM} & =\frac{a(1+\delta)}{\sqrt{2-2\text{cos}(\theta_b)}},~~~\mathbf{b_1}^{SM}=\mathbf{b}^{SM}[0,1,0,1]\\
		\lambda_2^{SM} & =\frac{a(1+\delta)}{\sqrt{(2-\delta)(1-\text{cos}(\theta_b))+\delta^2-\sqrt{3}\delta \text{sin}(\theta_b)}},~~~\mathbf{b_2}^{SM}=\mathbf{b}^{SM}[0,1,-1,-1]\\
		\lambda_3^{SM} & =\frac{a(1+\delta)}{\sqrt{2+3\delta^2-2\text{cos}(\theta_b)-2\sqrt{3}\delta \text{sin}(\theta_b)}},~~~\mathbf{b_3}^{SM}=\mathbf{b}^{SM}[0,1,-2,0]\\
		\lambda_4^{SM} & =\frac{a(1+\delta)}{\sqrt{(2+\delta)(1-cos(\theta_b))+\delta^2+\sqrt{3}\delta \text{sin}(\theta_b)}}, ~~~\mathbf{b_3}^{SM}=\mathbf{b}^{SM}[1,0,0,-1]
	\end{split}
\end{equation}

%%%%%%%%%%%%%%%%%%  FIGURE 2 %%%%%%%%%%%%%%%%%%%%%%%%%%%%%%%%%%

\begin{figure*}[tbh]
	\includegraphics[width=0.4\textwidth]{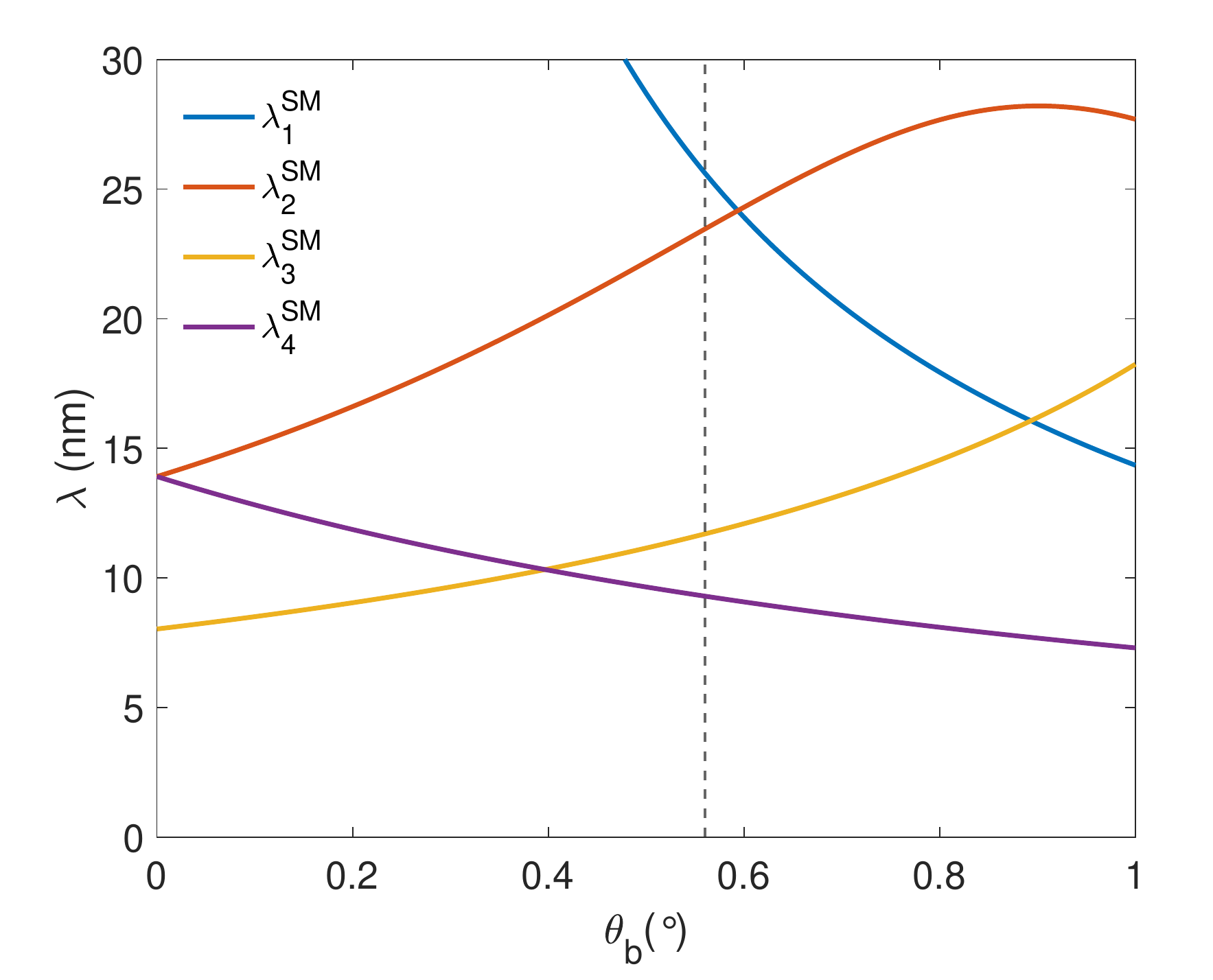}
	\caption{Evolution of supermoir\'e wavelength with $\theta_b$, for $\theta_t\sim 0^{\circ}$. In our case $\theta_b\sim 0.56^{\circ}$, shown by black dashed line. Using Eq.(S6) gives $\lambda_r^{SM}\sim 25.6,~23.4,~11.7 ~\text{and}~9.3$nm.}
	\label{fig:fig2}
\end{figure*}
%%%%%%%%%%%%%%%%%%%%%%%%%%%%%%%%%%%%%%%%%%%%%%%%%%%%%%%%%%%%%%%%%%%%%%%%%
\newpage

\section{Temperature dependence for the electron side}
$\rho-$T for the electron side in between SDP1 and SDP2. The resistivity scales linearly with temperature with a maximum slope of $d\rho/dT=A=0.9~\Omega$/K.
\begin{figure*}[tbh]
	\includegraphics[width=0.4\textwidth]{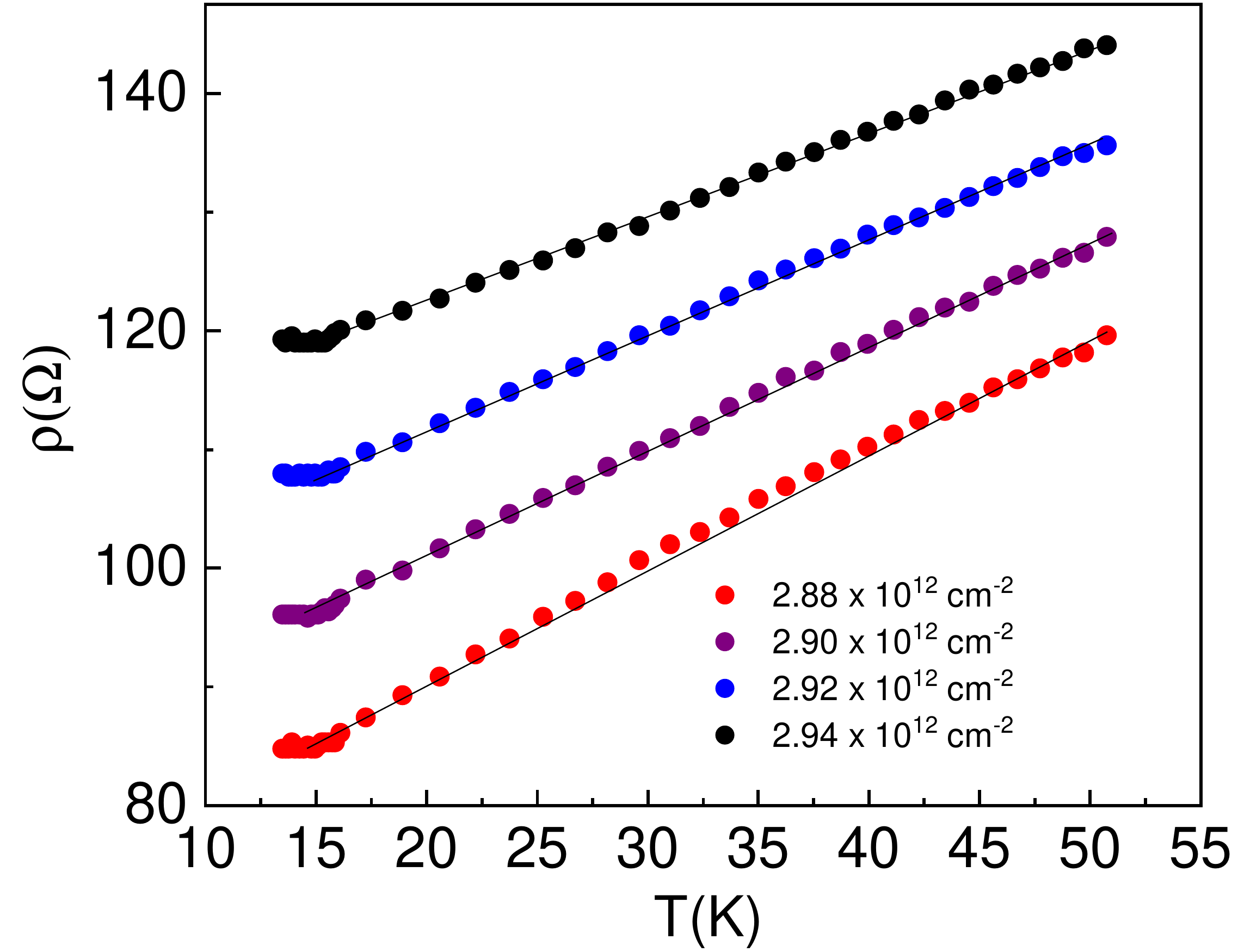}
	\caption{ Temperature dependance in the region $n_{s1}<n<n_{s1}$, for the electron doped region. The curves has been offset with 10$\Omega$ for clarity. The data was fitted in with $\rho\sim \rho_0+AT$.}
	\label{fig:fig2}
\end{figure*}
%%%%%%%%%%%%%%%%%%%%%%%%%%%%%%%%%%%%%%%%%%%%%%%%%%%%%%%%%%%%%%%%%%%%%%%%%
%\newpage
%%%%%%%%%%%%%%%%%%%%%%figure 1%%%%%%%%%%%%%%%%%%%%%%%%%%%
%\begin{figure*}[ht!]
% \includegraphics[width=0.8\textwidth]{sp_fig1.pdf}
% \caption{(Color Online) (a) Schematic of the equivalent capacitive circuit of a dual gated BLG device. $C_p~^b$ ($C_p~^t$) are the parasitic capacitances. (b) Simplified equivalent circuit. $C_t$ is the measured total capacitance. (c) Extraction procedure to convert gate voltages ($V_{bg}, V_{bg}$) to Fermi energy ($E_F$) along constant $\bar D$ line. }
% \label{fig:sp1}
%\end{figure*}
%%%%%%%%%%%%%%%%%%%%%%%%%%%%%%%%%%%%%%%%%%%%%%%%%%%%%%%%%%
%

%\bibliography{ref_sblg}{}

\end{document}